\documentclass[sigconf]{acmart}
\AtBeginDocument{%
  }

\setcopyright{acmlicensed}
\copyrightyear{2018}
\acmYear{2018}
\acmDOI{XXXXXXX.XXXXXXX}
\acmConference[Conference acronym 'XX]{Make sure to enter the correct
  conference title from your rights confirmation email}{June 03--05,
  2018}{Woodstock, NY}
\acmISBN{978-1-4503-XXXX-X/2018/06}

\usepackage{booktabs}
\usepackage{multirow}
\usepackage{geometry}
\usepackage{caption}

\copyrightyear{2026}
\acmYear{2026}
\setcopyright{cc}
\setcctype{by}
\acmConference[CHI EA '26]{Extended Abstracts of the 2026 CHI Conference on Human Factors in Computing Systems}{April 13--17, 2026}{Barcelona, Spain}
\acmBooktitle{Extended Abstracts of the 2026 CHI Conference on Human Factors in Computing Systems (CHI EA '26), April 13--17, 2026, Barcelona, Spain}
\acmDOI{10.1145/3772363.3798395}
\acmISBN{979-8-4007-2281-3/2026/04}

\begin{document}

\title{Auto-Generating Personas from User Reviews in VR App Stores}


\author{Yi Wang}
\affiliation{%
 \institution{School of Information Technology, Deakin University}
 \city{Melbourne}
 \country{Australia}}
 \email{xve@deakin.edu.au}

\author{Kexin Cheng}
\affiliation{%
  \institution{School of Information Technology, Deakin University}
 \city{Melbourne}
 \country{Australia}}
 \email{s224384854@deakin.edu.au}

 \author{Xiao Liu}
\affiliation{%
  \institution{School of Information Technology, Deakin University}
 \city{Melbourne}
 \country{Australia}}
 \email{xiao.liu@deakin.edu.au}

\author{Chetan Arora}
\affiliation{%
  \institution{Faculty of Information Technology, Monash University}
 \city{Melbourne}
 \country{Australia}}
  \email{chetan.arora@monash.edu}

  \author{John Grundy}
\affiliation{%
  \institution{Faculty of Information Technology, Monash University}
 \city{Melbourne}
 \country{Australia}}
  \email{john.grundy@monash.edu}

\author{Thuong Hoang}
\affiliation{%
  \institution{School of Information Technology, Deakin University}
 \city{Melbourne}
 \country{Australia}}
\email{thuong.hoang@deakin.edu.au}

  \author{Henry Been-Lirn Duh}
\affiliation{%
  \institution{School of Design, The Hong Kong Polytechnic University}
 \city{Hong Kong}
 \country{China}}
  \email{henry.duh@polyu.edu.hk}
\renewcommand{\shortauthors}{Trovato et al.}

\begin{abstract}
  Personas are a valuable tool for discussing accessibility requirements in software design and development practices. However, the use of personas for accessibility-focused requirements elicitation in VR projects remains limited and is accompanied by several challenges. To fill this gap, we developed an auto-generated persona system in a VR course, where the personas were used to facilitate discussions on accessibility requirements and to guide VR design and development. Our findings indicate that the auto-generated persona system enabled students to develop empathy more efficiently. This study demonstrates the use of automatically generated personas in VR course settings as a means of eliciting latent accessibility requirements.
  
\end{abstract}


\maketitle

\section{Introduction}



Personas are an important requirements tool in user-centered design (UCD) \cite{Salminen22, dantin2005application} and requirements engineering \cite{WANG2025107609, billestrup2014persona}. Traditional approaches to creating personas primarily rely on data from qualitative and quantitative methods, including surveys, interviews, online data, and observations \cite{salminen2020literature, farooq2025representing}. Some personas are also constructed from large-scale datasets, such as social media data \cite{jung2017persona} and customer data \cite{mcginn2008data}. However, obtaining large-scale datasets poses significant challenges and raises ethical concerns, particularly for novices or students without sufficient technical expertise.

With the rapid development of large language models (LLMs) in recent years, an increasing number of studies have have proposed LLM-based systems for automated persona generation \cite{Zhang24, Choi2025, Sun25}. Recent work suggested that LLMs can significantly reduce the time and effort required to generate personas and can rapidly produce diverse personas \cite{ainen23}. Consequently, some studies have introduced LLM-based persona generation tools into undergraduate education, such as UCD course practices \cite{Zhang24}, where students often lack sufficient expertise in data analysis and may otherwise construct superficial or fabricated personas.

Beyond persona generation, LLMs are applied to enhance the accessibility of software projects, such as web accessibility \cite{pedemonte2025improving}. LLMs have also been used to raise students’ awareness of accessibility issues in software projects \cite{aljedaani2025enhancing}. Meanwhile, personas are frequently employed in accessibility testing \cite{henka2014persona}. A recent study further explored the use of LLMs to construct personas representing individuals with down syndrome \cite{Sun25}. Furthermore, accessibility challenges in virtual reality (VR) are fundamentally different from those in traditional desktop or mobile applications \cite{Wang2025VR}. For example, VR systems introduce distinct interaction constraints, such as motion sickness and limitations in spatial navigation. However, in recent years there has been a lack of innovative approaches to addressing VR-specific accessibility challenges in early-stage design education.




In this paper, we developed a web-based persona generation system that automatically captures user requirements from the Meta and Steam VR stores and extracts accessibility-related user reviews. Based on these reviews, the system automatically constructs personas with accessibility needs and summarizes the accessibility requirements they contain. The system supports conversational interaction: users can provide their project type and description, and the system automatically matches the most prominent accessibility-related reviews within that type to generate personas. Users may also request recommendations for personas related to specific requirements or personas constructed from reviews of the same disability type in other VR applications. To mitigate hallucinations commonly associated with LLMs, the system integrates an LLM (GPT-4o) with a Retrieval-Augmented Generation (RAG) framework, an approach that has previously been shown to outperform alternatives \cite{Sun25}. In this study, the goal of persona generation is not the artifact itself, but its function in (1) anchoring accessibility discussions in real user evidence, (2) mitigating abstraction during early VR requirements elicitation, and (3) facilitating perspective-taking in educational design contexts. To that end, our study is focused on answering the following research question (RQ):

\textbf{RQ. }To what extent do automatically generated accessibility personas impact students’ empathy?


We found that the system supported empathy in students’ discussions of accessibility requirements, enabling them to consider accessibility from the perspective of users with disabilities and enhanced students’ sense of design responsibility. To our knowledge, this is the first integration of automatically generated accessibility personas into VR course instruction, demonstrating the potential to advance inclusive VR education.

\section{Related Work}

Personas are an essential tool in UCD practice, helping design teams discuss and identify user needs in a vivid and engaging way \cite{nielsen2013personas}. In requirements engineering, personas are commonly employed to capture and articulate user requirements, while also addressing potential issues such as overlooked accessibility needs and challenges in team collaboration and communication \cite{WANG2025107609}. However, the use of personas in requirements engineering still faces many challenges and limitations, particularly in immersive systems such as VR \cite{WANG2025107609}. Meanwhile, VR development often follows game design workflows and informal practices. Traditional requirements engineering methods, such as elicitation, analysis, and specification, may lead to information loss and increased costs \cite{Karre2024}. Schneidewind et al. \cite{Schneidewind2012} found that integrating personas into requirements engineering processes can help model, prioritize, and validate requirements more effectively. However, it remains unclear how personas can be systematically leveraged as a tool to elicit user requirements in VR projects.



Personas have been widely adopted in various educational domains. For example, Silva and Motti used personas in higher education and found that persona-based empathy strengthened students’ ability to understand and apply accessibility principles \cite{Silva2024}. Prior work has also demonstrated that personas can foster empathy in vocational education \cite{van2012based}. In recent years, an increasing number of studies have started to investigate the use of LLMs to build personas in educational practices. For example, Zhang et al. \cite{zhang2023personagen, Zhang24} leveraged LLMs to automatically generate personas from students’ survey data. This method also helped prevent situations in which students lacking data analysis skills might generate fictional personas \cite{Sun25}. Furthermore, Wang et al. \cite{wang2026discussingneedsvrnovel} proposed a new method for embedding personas within VR environments to support discussions of accessibility requirements, which was found to enhance participants’ perceived social presence. Henka and Zimmermann \cite{Henka14} introduced a novel method for mapping WCAG (Web Content Accessibility Guidelines) to personas, thereby facilitating developers’ understanding of accessibility issues in user-centered contexts. Loitsch et al. \cite{Loitsch16} noted that despite the availability of various accessibility tools, frameworks, and detection systems, meaningful accessibility relies fundamentally on knowledge and awareness. They therefore used personas as a teaching medium and demonstrated their effectiveness in supporting accessibility education. Although automated persona creation tools have been applied in various educational settings, little is known about how they can be integrated into VR courses to help students more effectively discuss and understand accessibility requirements and address accessibility issues in their VR projects.

To fill this gap, we introduce a web-based persona generation system that leverages LLMs and a Retrieval-Augmented Generation (RAG) framework to automatically generate personas grounded in accessibility-related user reviews. To our knowledge, this work is among the first to systematically leverage VR store user reviews as a data source for persona generation in VR design education.




\section{System Overview}

\begin{figure*}
    \centering
    \includegraphics[width=1\linewidth]{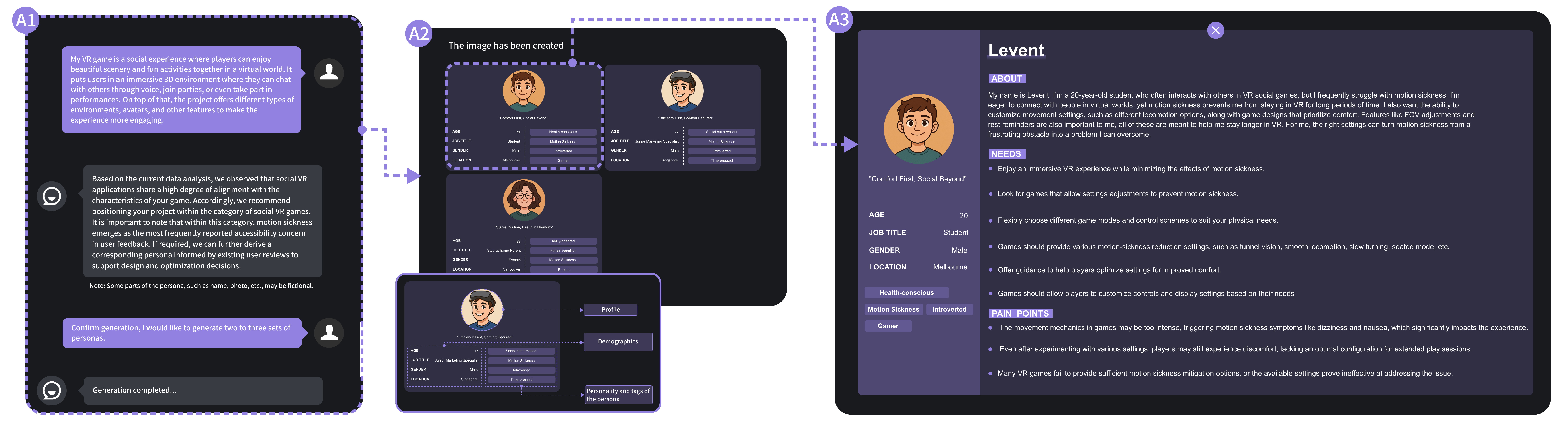}
    \caption{Features of the system: \textbf{A1} represents the dialogue between students and the system. \textbf{A2} and \textbf{A3} refer the automatically generated persona thumbnail and the detailed persona description, respectively. }
    \label{fig:2}
\end{figure*}

Based on these design goals, we present the system (see Figure \ref{fig:2}), which supports students in VR courses by enabling them to support more systematic understanding of accessibility requirements in VR through conversational interaction. The system integrates Large Language Models (LLMs) within a Retrieval-Augmented Generation (RAG) framework to efficiently generate personas grounded in accessibility-related user reviews. The system was developed using React for the web application frontend and Python for the backend.




\subsection{System Features and Implementation }

\subsubsection{Data Source and Processing. }We limited data collection to the 50 most popular VR applications. Prior work \cite{wang2025investigating} indicated that less popular or lower-rated VR applications contain very limited user reviews, which may hinder the system's ability to generate comprehensive and well-grounded persona templates. Since the Meta Quest Store does not provide a public application programming interface (API), we employed Web scraping techniques to collect user review data. In contrast, Steam provides a public API that allows for the direct extraction of user review data and tags.

During data scraping, we employed predefined disability-related keywords and fuzzy matching to improve the precision of accessibility-related review identification, drawing on World Health Organization (WHO) classifications and prior work \cite{wang2025investigating}. This approach ensured that accessibility-related reviews were aligned with internationally recognized disability domains rather than ad hoc keyword clusters. Reviews shorter than 20 words were removed, as longer reviews are more likely to contain user characteristics \cite{Choi2025}. We further excluded advertisements, non-English reviews, and content containing insults or discriminatory language based on issues identified during manual review by two researchers. After reprocessing, two researchers reviewed VR application names, tags, and official descriptions to determine each application’s primary category, as many titles contained multiple VR-related tags (e.g., action, horror, multiplayer). All reviews were then categorized by VR application type, including action, social, horror, puzzle, simulation, and sports. Accessibility-related reviews from the same category (e.g., motion sickness reports in action VR applications across Meta and Steam) were consolidated, resulting in 396 high-quality reviews. Finally, the cleaned reviews were segmented into semantically coherent chunks and embedded using a sentence-transformer model. The embeddings and metadata were stored in the Chroma vector database, enabling efficient semantic retrieval during persona generation and ensuring that generated personas were grounded in authentic user feedback rather than keyword matching alone.

\subsubsection{Retrieval-Augmented Persona Generation. }We adopted a RAG framework, as prior work has shown that RAG can improve the grounding of large language model outputs in retrieved evidence \cite{Sun25, ram2023context}. The system first queried the vector database to retrieve the top semantically relevant review segments based on the selected VR type and disability group. These evidence chunks are then injected into the GPT-4o prompt, where GPT-4o generates an intermediate user summary and extracts structured dimension–value pairs. For example, \textit{dimensions} represent mutually exclusive disability categories (e.g., motion sickness, hearing loss), while \textit{values} represent accessibility requirements, pain points, and demographic information. This intermediate representation constrains generation by organizing evidence into structured components, thereby supporting coherence and mitigating hallucination risks. Finally, the dimension–value pairs are compiled into a standardized persona containing a brief biography, pain points, representative quotes directly grounded in user reviews, and explicit requirements. Profile photos were generated using DALL·E 3 based on demographic information.

\section{User Study}


\subsection{Participants}

We recruited a class of 24 students (10 males and 14 females), all of whom had completed a user-centered design course and were familiar with the use of personas and methods for analyzing user needs. Their ages ranged from 22 to 24 (\textit{M} = 22.4, \textit{SD} = 1.22). All students were required to be enrolled in a VR course. Participants were recruited from a single class by contacting the course instructor.

\subsection{Procedure}

We conducted a two-week, face-to-face teaching program comprising a maximum of 8 hours of in-person instruction. 

In the first week, the session introduced core concepts of accessibility and requirements through illustrative cases, including challenges encountered by users with visual and hearing impairments. Students then participated in group discussions to share their understanding of accessibility requirements and related examples. This activity lasted approximately two hours, including a short break. The session concluded with an introduction to VR accessibility, covering its definition, importance, existing standards, and key dimensions (e.g., visual, auditory, and motor accessibility). Students were finally asked to reflect on their own VR project types and identify potential accessibility issues.

In the second session, the researchers introduced the system and the survey-based approach. This session lasted 20 minutes. The researchers then randomly assigned the students into two groups: one group used the system to create personas, while the other group employed the survey-based approach. Groups in the system condition interacted with the tool by describing their VR project types and project contexts. The system automatically retrieved relevant data and highlighted the most prevalent accessibility-related reviews for the specified category. Students could further query accessibility requirements summarized in the generated personas and request comparable personas across different VR applications or disability types. In contrast, the control group created personas based on prior UCD course practices and supplemented them with materials they independently collected from online resources, academic literature, forums, and VR store reviews. Unlike the system condition, they were not provided with pre-filtered accessibility-related content, and persona creation was limited to a one-hour session. Subsequently, one researcher and two instructors met with each group to discuss their understanding of the identified accessibility requirements and assess whether students could accurately interpret these requirements through the personas. During discussions, groups could consult online resources, examine existing VR applications, and experience selected examples using Meta Quest 3S headsets. Students were also encouraged to produce low-fidelity sketches and discuss potential future design. The discussion phase lasted up to one hour.

In the third session, groups switched conditions: the system group adopted the survey-based approach, while the survey group used the system, and completing the same tasks as in the previous session. After a 15-minute break, all participants completed a post-study questionnaire, followed by face-to-face semi-structured group interviews in which students reflected on both their own perspectives and those of others.

\section{Measures and Analysis}

Under both conditions, participants completed questionnaires assessing empathy using a 7-point Likert scale. Empathy was measured using selected subscales of the Interpersonal Reactivity Index (IRI) \cite{de2007measuring}. Specifically, we included three subscales relevant to this study: perspective taking, empathic concern, and fantasy. Prior to inferential analysis, the normality of the quantitative data was assessed using the Shapiro–Wilk test. As the assumption of normality was satisfied, paired-samples t-tests were conducted to examine differences in empathy scores between the two conditions \cite{ross2017paired}.

To supplement the quantitative data, we conducted interviews in small groups of two to three students to to explore how the system influenced students’ empathy development and to collect suggestions for future improvements. Interviews were audio-recorded and transcribed verbatim. Thematic analysis was conducted on the qualitative data using MAXQDA. Three researchers independently coded the data after familiarization. Following each round, the codes were discussed and consolidated with reference to the research questions until data saturation was reached, and a final codebook was produced \cite{braun2006using}. This iterative discussion and consolidation process was conducted over three rounds.



\section{Results}


The students were reported higher scores for the system in terms of perspective taking, empathic concern, and fantasy. Specifically, the system scored higher than the survey-based approach (\textit{t} = 2.989, \textit{p} = .015, system: \textit{M} = 4.45, \textit{SD} = 0.78, survey-based approach: \textit{M} = 3.06, \textit{SD} = 1.39).


\textbf{Perspective Taking. }The students were reported higher scores in the system condition compared to the survey-based condition (\textit{t} = 3.715, \textit{p} = .004, system: \textit{M} = 4.65, \textit{SD} = 0.81, survey-based approach: \textit{M} = 3.25, \textit{SD} = 1.24). Most students reported that requirement discussions based on accessibility personas enabled them to gain a deeper understanding of the accessibility challenges encountered in VR usage. Some students described the experience as unprecedented, acknowledging that they had not previously recognized the impact of accessibility issues on people with disabilities and had tended to regard VR merely as a novel technology (P10, P21, P22, P24). Furthermore, P14 reported \textit{``Now I’ll naturally start asking myself, can people with disabilities use this feature? And if not, is there an alternative?''}

\textbf{Empathic Concern. }The students were reported higher scores in the system condition compared to the survey-based condition (\textit{t} = 2.515, \textit{p} = .033, system: \textit{M} = 4.35, \textit{SD} = 1.29, survey-based approach: \textit{M} = 2.85, \textit{SD} = 1.54). Many students reported that after using the system, they became more attentive to their emotions and their need to be treated equally. Some students noted that this approach motivated them to address accessibility issues more actively (P5, P10, P11, P18, P21). Further, P4 indicated \textit{``He [persona] both liked it and felt helpless about it, which gave me a deeper sense of his mixed feelings.''} P6 stated \textit{``It makes me wonder if I’ve also been part of creating this unfair experience.''} P16 indicated their viewpoint \textit{``I realized how much frustration users with motor difficulties feel when they face complex interactions… and honestly, that kind of emotional experience is something I rarely paid attention to before.''} Therefore, empathic concern may influence students’ emotions, such as triggering negative personal feelings.

\textbf{Fantasy. }No significant difference was observed between the system and the survey-based approach (system: \textit{M} = 4.15, \textit{SD} = 2.90, survey-based approach: \textit{M} = 3.10, \textit{SD} = 1.96). However, some students indicated that the generated accessibility personas enabled them to better empathize with the experiences of people with disabilities. For example, P12 provided an example \textit{``There was one persona who couldn’t use the controllers but still longed to walk freely in VR...it was the first time I felt this wasn’t just made up, there are really people out there facing this situation.''}

\section{Discussion}

Empathy is a critical component in understanding inclusivity and accessibility \cite{Bennett19, horton2021empathy}. Our results revealed that the system supported empathy development in the context of VR course practices. Specifically, most students reported that they were able to adopt the perspective of users with disabilities, and the system also reduced the perceived abstraction or detachment often associated with fictional personas. Moreover, although our primary focus was on accessibility requirements, the automated creation of personas based on real user data further supported students in developing a deeper understanding of user needs \cite{Zhang24}. Many students reported that they had rarely considered accessibility requirements and often overlooked the needs of people with disabilities.

Many participants reported surprise when first encountering the accessibility challenges represented in the personas, as they were unaware that even mature VR applications still face substantial accessibility issues. Previous work similarly reported that free VR applications often contain numerous accessibility barriers \cite{Naikar2024}. Some students also described experiences of self-reflection, indicating that the system appeared to stimulate ethical reflection. In this respect, the system functioned as a tool for evoking empathic concern and encouraging students to consider for the needs of people with disabilities in VR design and development. Meanwhile, although the system adopts a conversational interface, our findings do not suggest that conversational interaction is inherently more effective than other modalities. Rather, the primary contribution of this study concerns whether persona construction grounded in structured real-user data can facilitate improvements in participants’ empathic understanding.

Although no statistically significant differences were observed in the fantasy subscale between the system and survey-based conditions, some students perceived the current presentation as insufficient to meaningfully enhance imaginative engagement. Participants suggested extending the system with scenario-based simulations and multimodal representations, such as visualized maps of accessibility pain points, interactive scenario simulations, and VR-based explorations of accessibility requirements \cite{Wang-ISMAR22}. Prior research has shown that first-person VR simulations of disability-related constraints can reduce both implicit and explicit biases \cite{Jang25}. Building on this insight, further enhancing the system’s interactive and experiential components may promote deeper imaginative engagement and empathy toward the needs of people with disabilities.

\subsection{Limitations and Future Work}

Although the system introduces an innovative approach that leverages LLMs and the RAG framework for accessibility-focused persona generation in VR courses, several limitations should be acknowledged.

At present, we evaluated the system with only 24 undergraduate students, who do not represent the broader population of learners. Moreover, students engaged with the system for only a limited period, which may have hindered their full involvement. In the future, we plan to deploy the system across an entire VR course and ensure that the accessibility requirements identified by students are meaningfully incorporated into their projects rather than being recognized but subsequently neglected in later course practices.

Additionally, although our system was associated with increased perceived empathic understanding of users with disabilities, empathy is not inherently or exclusively beneficial. As emphasized by Bennett et al. \cite{Bennett19}, empathy may lead to emotional over-identification (``feeling like''), which can in turn result in misinterpretations of users’ actual needs. Therefore, we plan to incorporate structured reflective prompts in future work to mitigate the risk of misguided empathy.

A noteworthy nuance is that students achieved lower scores in the survey-based condition, even though they were tasked with independently gathering data and constructing accessibility personas. Although this method may promote deeper involvement, it also entails limitations, as students might depend on self-reports or poorly designed survey instruments that may not sufficiently capture the lived realities of accessibility challenges. Future research should explicitly examine this difference by comparing system-supported methods with direct engagement approaches, such as conducting interviews with people with disabilities.

From a technical perspective, although LLMs integrated with a RAG framework is considered effective in reducing hallucinations, this approach still struggles to eliminate stereotypical influences in accessibility-related reviews \cite{Sun25}. Therefore, the lack of assessment of these stereotypes is a significant limitation of our work. Future research should investigate the potential stereotypes embedded in the personas generated by the system. In addition, we did not evaluate whether students’ implicit and explicit biases toward people with disabilities were reduced, primarily because our study was conducted in a course setting where the main objective was to effectively identify and discuss accessibility requirements in VR. Future research should specifically recruit a broader or more diverse sample to examine changes in bias before and after using the system. Furthermore, we plan to evaluate the system’s user experience, usability, and workload.



\section{Conclusion}

The auto-generated persona system presents a structured approach to constructing accessibility-oriented personas using LLMs within a RAG framework, deployed in the context of VR design education. This approach aims to help students discuss and understand accessibility requirements in VR more efficiently and accurately. We deployed the system in a VR course. The results showed that the system was associated with increased empathy. Finally, we discuss the potential limitations of our work and propose several directions for future research.


\bibliographystyle{ACM-Reference-Format}
\bibliography{sample-base}

@String{Computing = "Computing" }

@String{Computer = "{IEEE} Computer" }

@String{Springer = "Springer-Verlag" }

@article{WANG2025107609,
title = {Who uses personas in requirements engineering: The practitioners’ perspective},
journal = {Information and Software Technology},
volume = {178},
pages = {107609},
year = {2025},
issn = {0950-5849},
doi = {https://doi.org/10.1016/j.infsof.2024.107609},
url = {https://www.sciencedirect.com/science/article/pii/S0950584924002143},
author = {Yi Wang and Chetan Arora and Xiao Liu and Thuong Hoang and Vasudha Malhotra and Ben Cheng and John Grundy},

}

@article{wang2025investigating,
  title={Investigating VR Accessibility Reviews for Users with Disabilities: A Qualitative Analysis},
  author={Wang, Yi and Arora, Chetan and Liu, Xiao and Hoang, Thuong and Zhang, Zhengxin and Duh, Henry Been Lirn and Grundy, John},
  journal={arXiv preprint arXiv:2508.13051},
  year={2025}
}

@inproceedings{zhang2023personagen,
  title={Personagen: A tool for generating personas from user feedback},
  author={Zhang, Xishuo and Liu, Lin and Wang, Yi and Liu, Xiao and Wang, Hailong and Ren, Anqi and Arora, Chetan},
  booktitle={2023 IEEE 31st International Requirements Engineering Conference (RE)},
  pages={353--354},
  year={2023},
  organization={IEEE}
}

@inproceedings{Zhang24,
author = {Zhang, Xishuo and Liu, Lin and Wang, Yi and Liu, Xiao and Wang, Hailong and Arora, Chetan and Liu, Haichao and Wang, Weijia and Hoang, Thuong},
title = {Auto-Generated Personas: Enhancing User-centered Design Practices among University Students},
year = {2024},
isbn = {9798400703317},
publisher = {Association for Computing Machinery},
address = {New York, NY, USA},
url = {https://doi.org/10.1145/3613905.3651043},
doi = {10.1145/3613905.3651043},
booktitle = {Extended Abstracts of the CHI Conference on Human Factors in Computing Systems},
articleno = {52},
numpages = {7},
location = {Honolulu, HI, USA},
series = {CHI EA '24}
}

@inproceedings{Salminen22,
author = {Salminen, Joni and Wenyun Guan, Kathleen and Jung, Soon-Gyo and Jansen, Bernard},
title = {Use Cases for Design Personas: A Systematic Review and New Frontiers},
year = {2022},
isbn = {9781450391573},
publisher = {Association for Computing Machinery},
address = {New York, NY, USA},
url = {https://doi.org/10.1145/3491102.3517589},
doi = {10.1145/3491102.3517589},
booktitle = {Proceedings of the 2022 CHI Conference on Human Factors in Computing Systems},
articleno = {543},
numpages = {21},
location = {New Orleans, LA, USA},
series = {CHI '22}
}

@article{de2007measuring,
  title={Measuring empathic tendencies: Reliability and validity of the Dutch version of the Interpersonal Reactivity Index},
  author={De Corte, Kim and Buysse, Ann and Verhofstadt, Lesley L and Roeyers, Herbert and Ponnet, Koen and Davis, Mark H},
  journal={Psychologica Belgica},
  volume={47},
  number={4},
  year={2007}
}

@article{braun2006using,
  title={Using thematic analysis in psychology},
  author={Braun, Virginia and Clarke, Victoria},
  journal={Qualitative research in psychology},
  volume={3},
  number={2},
  pages={77--101},
  year={2006},
  publisher={Taylor \& Francis}
}

@incollection{ross2017paired,
  title={Paired samples T-test},
  author={Ross, Amanda and Willson, Victor L},
  booktitle={Basic and advanced statistical tests: Writing results sections and creating tables and figures},
  pages={17--19},
  year={2017},
  publisher={Springer}
}

@inproceedings{Choi2025,
author = {Choi, Yoonseo and Kang, Eun Jeong and Choi, Seulgi and Lee, Min Kyung and Kim, Juho},
title = {Proxona: Supporting Creators' Sensemaking and Ideation with LLM-Powered Audience Personas},
year = {2025},
isbn = {9798400713941},
publisher = {Association for Computing Machinery},
address = {New York, NY, USA},
url = {https://doi.org/10.1145/3706598.3714034},
doi = {10.1145/3706598.3714034},
booktitle = {Proceedings of the 2025 CHI Conference on Human Factors in Computing Systems},
articleno = {149},
numpages = {32},
location = {
},
series = {CHI '25}
}

@inproceedings{Sun25,
author = {Sun, Lipeipei and Qin, Tianzi and Hu, Anran and Zhang, Jiale and Lin, Shuojia and Chen, Jianyan and Ali, Mona and Prpa, Mirjana},
title = {Persona-L has Entered the Chat: Leveraging LLMs and Ability-based Framework for Personas of People with Complex Needs},
year = {2025},
isbn = {9798400713941},
publisher = {Association for Computing Machinery},
address = {New York, NY, USA},
url = {https://doi.org/10.1145/3706598.3713445},
doi = {10.1145/3706598.3713445},
booktitle = {Proceedings of the 2025 CHI Conference on Human Factors in Computing Systems},
articleno = {1109},
numpages = {31},
location = {
},
series = {CHI '25}
}

@article{ram2023context,
  title={In-context retrieval-augmented language models},
  author={Ram, Ori and Levine, Yoav and Dalmedigos, Itay and Muhlgay, Dor and Shashua, Amnon and Leyton-Brown, Kevin and Shoham, Yoav},
  journal={Transactions of the Association for Computational Linguistics},
  volume={11},
  pages={1316--1331},
  year={2023},
  publisher={MIT Press One Broadway, 12th Floor, Cambridge, Massachusetts 02142, USA~…}
}

@inproceedings{billestrup2014persona,
  title={Persona usage in software development: advantages and obstacles},
  author={Billestrup, Jane and Stage, Jan and Nielsen, Lene and Hansen, Kira S},
  booktitle={The Seventh International Conference on Advances in Computer-Human Interactions, ACHI},
  pages={359--364},
  year={2014}
}

@inproceedings{dantin2005application,
  title={Application of personas in user interface design for educational software},
  author={Dantin, Ursula},
  booktitle={Proceedings of the 7th Australasian conference on Computing education-Volume 42},
  pages={239--247},
  year={2005}
}

@inproceedings{Bennett19,
author = {Bennett, Cynthia L. and Rosner, Daniela K.},
title = {The Promise of Empathy: Design, Disability, and Knowing the "Other"},
year = {2019},
isbn = {9781450359702},
publisher = {Association for Computing Machinery},
address = {New York, NY, USA},
url = {https://doi.org/10.1145/3290605.3300528},
doi = {10.1145/3290605.3300528},
booktitle = {Proceedings of the 2019 CHI Conference on Human Factors in Computing Systems},
pages = {1–13},
numpages = {13},
location = {Glasgow, Scotland Uk},
series = {CHI '19}
}

@article{horton2021empathy,
  title={Empathy cannot sustain action in technology accessibility},
  author={Horton, Sarah},
  journal={Frontiers in Computer Science},
  volume={3},
  pages={617044},
  year={2021},
  publisher={Frontiers Media SA}
}

@inproceedings{Naikar2024,
author = {Naikar, Vinaya Hanumant and Subramanian, Shwetha and Tigwell, Garreth W.},
title = {Accessibility Feature Implementation Within Free VR Experiences},
year = {2024},
isbn = {9798400703317},
publisher = {Association for Computing Machinery},
address = {New York, NY, USA},
url = {https://doi.org/10.1145/3613905.3650935},
doi = {10.1145/3613905.3650935},

booktitle = {Extended Abstracts of the CHI Conference on Human Factors in Computing Systems},
articleno = {31},
numpages = {9},
location = {Honolulu, HI, USA},
series = {CHI EA '24}
}

@ARTICLE{Jang25,
  author={Jang, Hyuckjin and Lee, Jeongmi},
  journal={IEEE Transactions on Visualization and Computer Graphics}, 
  title={Effective VR Intervention to Reduce Implicit Bias Towards People with Physical Disabilities: The Interplay Between Experience Design and Individual Characteristics}, 
  year={2025},
  volume={31},
  number={5},
  pages={2342-2352},
  doi={10.1109/TVCG.2025.3549532}}

@INPROCEEDINGS{Wang-ISMAR22,
  author={Wang, Yi and Liu, Xiao and Cheng, Ben and Arora, Chetan and Hoang, Thuong},
  booktitle={2022 IEEE International Symposium on Mixed and Augmented Reality Adjunct (ISMAR-Adjunct)}, 
  title={VR4HcRE: Virtual Reality Platform for Human-centric Requirements Elicitation}, 
  year={2022},
  volume={},
  number={},
  pages={788-793},
  doi={10.1109/ISMAR-Adjunct57072.2022.00168}}

@book{nielsen2013personas,
  title={Personas-user focused design},
  author={Nielsen, Lene},
  volume={15},
  year={2013},
  publisher={Springer}
}

@inproceedings{Wang2025VR,
  author    = {Wang, Yi and Liu, Xiao and Arora, Chetan and Grundy, John and Hoang, Thuong},
  title     = {Understanding VR Accessibility Practices of VR Professionals},
  booktitle = {Proceedings of the 2025 CHI Conference on Human Factors in Computing Systems},
  year      = {2025},
  pages     = {1--17}
}

@INPROCEEDINGS{Schneidewind2012,
  author={Schneidewind, Lydia and Hörold, Stephan and Mayas, Cindy and Krömker, Heidi and Falke, Sascha and Pucklitsch, Tony},
  booktitle={2012 First International Workshop on Usability and Accessibility Focused Requirements Engineering (UsARE)}, 
  title={How personas support requirements engineering}, 
  year={2012},
  volume={},
  number={},
  pages={1-5},
  doi={10.1109/UsARE.2012.6226786}}

@ARTICLE{Silva2024,
  author={Silva, Paula Alexandra and Motti, Vivian Genaro},
  journal={IEEE Access}, 
  title={Evaluating a Multi-Component Classroom Intervention to Teach Accessibility in Higher Education: A Case Study With Persona Cards}, 
  year={2024},
  volume={12},
  number={},
  pages={20299-20312},
  doi={10.1109/ACCESS.2024.3360135}}

@article{van2012based,
  title={based personas: Teaching empathy in professional education.},
  author={van Rooij, Shahron Williams},
  journal={Journal of Effective Teaching},
  volume={12},
  number={3},
  pages={77--86},
  year={2012},
  publisher={ERIC}
}

@inproceedings{salminen2020literature,
  title={A literature review of quantitative persona creation},
  author={Salminen, Joni and Guan, Kathleen and Jung, Soon-gyo and Chowdhury, Shammur A and Jansen, Bernard J},
  booktitle={Proceedings of the 2020 CHI conference on human factors in computing systems},
  pages={1--14},
  year={2020}
}

@article{farooq2025representing,
  title={Representing groups of students as personas: A systematic review of persona creation, application, and trends in the educational domain},
  author={Farooq, Ali and Alabed, Amani and Msefula, Pilira Stella and Tamime, Reham AL and Salminen, Joni and Jung, Soon-gyo and Jansen, Bernard J},
  journal={Computers and Education Open},
  pages={100242},
  year={2025},
  publisher={Elsevier}
}

@inproceedings{jung2017persona,
  title={Persona generation from aggregated social media data},
  author={Jung, Soon-Gyo and An, Jisun and Kwak, Haewoon and Ahmad, Moeed and Nielsen, Lene and Jansen, Bernard J},
  booktitle={Proceedings of the 2017 CHI conference extended abstracts on human factors in computing systems},
  pages={1748--1755},
  year={2017}
}

@inproceedings{mcginn2008data,
  title={Data-driven persona development},
  author={McGinn, Jennifer and Kotamraju, Nalini},
  booktitle={Proceedings of the SIGCHI conference on human factors in computing systems},
  pages={1521--1524},
  year={2008}
}

@inproceedings{ainen23,
author = {H\"{a}m\"{a}l\"{a}inen, Perttu and Tavast, Mikke and Kunnari, Anton},
title = {Evaluating Large Language Models in Generating Synthetic HCI Research Data: a Case Study},
year = {2023},
isbn = {9781450394215},
publisher = {Association for Computing Machinery},
address = {New York, NY, USA},
url = {https://doi.org/10.1145/3544548.3580688},
doi = {10.1145/3544548.3580688},

booktitle = {Proceedings of the 2023 CHI Conference on Human Factors in Computing Systems},
articleno = {433},
numpages = {19},
location = {Hamburg, Germany},
series = {CHI '23}
}

@article{pedemonte2025improving,
  title={Improving Web Accessibility with an LLM-Based Browser Extension: A Preliminary Evaluation},
  author={Pedemonte, Giacomo and Leotta, Maurizio and Ribaudo, Marina},
  journal={IEEE Access},
  year={2025},
  publisher={IEEE}
}

@inproceedings{aljedaani2025enhancing,
  title={Enhancing accessibility in software engineering projects with large language models (llms)},
  author={Aljedaani, Wajdi and Eler, Marcelo Medeiros and Parthasarathy, PD},
  booktitle={Proceedings of the 56th ACM Technical Symposium on Computer Science Education V. 1},
  pages={25--31},
  year={2025}
}

@inproceedings{henka2014persona,
  title={Persona Based Accessibility Testing: Towards User-Centered Accessibility Evaluation},
  author={Henka, Alexander and Zimmermann, Gottfried},
  booktitle={International Conference on Human-Computer Interaction},
  pages={226--231},
  year={2014},
  organization={Springer}
}

@article{Karre2024,
author = {Karre, Sai Anirudh and Reddy, Y. Raghu and Mittal, Raghav},
title = {RE Methods for Virtual Reality Software Product Development: A Mapping Study},
year = {2024},
issue_date = {May 2024},
publisher = {Association for Computing Machinery},
address = {New York, NY, USA},
volume = {33},
number = {4},
issn = {1049-331X},
url = {https://doi.org/10.1145/3649595},
doi = {10.1145/3649595},
journal = {ACM Trans. Softw. Eng. Methodol.},
month = apr,
articleno = {88},
numpages = {31}
}

@InProceedings{Loitsch16,
author="Loitsch, Claudia
and Weber, Gerhard
and Voegler, Jens",
editor="Miesenberger, Klaus
and B{\"u}hler, Christian
and Penaz, Petr",
title="Teaching Accessibility with Personas",
booktitle="Computers Helping People with Special Needs",
year="2016",
publisher="Springer International Publishing",
address="Cham",
pages="453--460",

isbn="978-3-319-41264-1"
}

@InProceedings{Henka14,
author="Henka, Alexander
and Zimmermann, Gottfried",
editor="Stephanidis, Constantine",
title="Persona Based Accessibility Testing",
booktitle="HCI International 2014 - Posters' Extended Abstracts",
year="2014",
publisher="Springer International Publishing",
address="Cham",
pages="226--231",

isbn="978-3-319-07854-0"
}

@misc{wang2026discussingneedsvrnovel,
      title={Discussing Your Needs in VR: A Novel Approach through Persona-based Stakeholder Role-Playing}, 
      author={Yi Wang and Zhengxin Zhang and Xiao Liu and Chetan Arora and John Grundy and Thuong Hoang},
      year={2026},
      eprint={2602.04632},
      archivePrefix={arXiv},
      primaryClass={cs.HC},
      url={https://arxiv.org/abs/2602.04632}, 
}


\end{document}